# ROI: A method for identifying organizations receiving personal data

David Rodriguez[1*], Jose M. Del Alamo[1*], Miguel Cozar[1] and Boni García[2]

[1*]ETSI Telecomunicación, Universidad Politécnica de Madrid, Spain.
[2]Campus Leganés, Universidad Carlos III de Madrid, Spain.

*Corresponding author(s). E-mail(s):
david.rtorrado@upm.es; jm.delalamo@upm.es;
Contributing authors: m.cozar@upm.es;
boni.garcia@uc3m.es;

**Abstract**

Many studies have exposed the massive collection of personal data in the digital ecosystem through, for instance, websites, mobile apps, or smart devices. This fact goes unnoticed by most users, who are also unaware that the collectors are sharing their personal data with many different organizations around the globe. This paper assesses techniques available in the state of the art to identify the organizations receiving this personal data. Based on our findings, we propose ROI (Receiver Organization Identifier), a fully automated method that combines different techniques to achieve a 95.71% precision score in identifying an organization receiving personal data. We demonstrate our method in the wild by evaluating 10,000 Android apps and exposing the organizations that receive users' personal data.

# 1 Introduction

The widespread adoption of fancy new smart devices, including many different sensors, facilitates the collection of personal data from individuals anywhere and anytime through the websites they visit and the apps they use. The distributed nature of the Internet further facilitates sharing these data with organizations worldwide [1].

Identifying the organizations that receive these personal data is becoming increasingly crucial for different stakeholders. For example, supervisory authorities may leverage this information to conduct investigations on the relationship between the source and destination of some personal data flows to understand a system's compliance with, for instance, legal requirements for international transfers of personal data [2]. Also, privacy and legal researchers can use this information to discover what companies are collecting massive amounts of personal data [3]. Additionally, app and web developers may want to check what organizations they send their users' personal data to, sometimes even without their knowledge [4], to meet transparency requirements set, e.g., by privacy regulations. Even app marketplaces can take advantage of it in their app review processes (e.g. [5], still in beta phase in June 2023) to help less experienced developers with their app regulatory compliance.

However, identifying the organizations receiving personal data is not an easy task. The app's or website's privacy policies, if present, often fail to include the third parties with which the collector is sharing the personal data [6]. Although a dynamic analysis of the collecting system and its network traffic can reveal the personal data flows [7] and the destination domains [8], identifying the organizations receiving the data may become challenging due to, e.g. WHOIS accuracy and reliability issues [9]. According to Libert et al. [10]: "we find that 36% of domains in our dataset have anonymous whois registration". We aim to advance the fundamental understanding of the domains receiving personal data flows and the organizations holding them. To this end, we have assessed two techniques available in the state of the art to identify the organization holding a domain, namely WHOIS service consultation and SSL certificate inspection. Our results show the performance of these individual techniques is far from desirable. Thus, we have developed a new technique based on the analyses of privacy policies and combined it into a new method (ROI - Receiver Organization Identifier), showing a high precision level (95.71%) in identifying the organization that receives personal data flows, and significantly outperforming similar methods available in the state of the art. Finally, to demonstrate its applicability in the wild, we have applied ROI to discover the companies receiving personal data on a sample of 10,000 Android apps.

Our original contributions are[1]:

1. A reliable and precise method to identify organizations holding domains that receive personal data flows, demonstrated in the wild in the Android ecosystem.
2. Two datasets supporting the validation of our method and the individual techniques, together with the assessment results. The first dataset includes

---

[1]The contributed datasets are available for review at https://drive.upm.es/s/tAwqgUj1s9KKuK1. They will be moved to an open data repository upon the paper acceptance.

142 privacy policies URLs annotated with the identity of the organization collecting the data. The second one consists of 300 domains and the organizations holding them.

3. An additional dataset of 1,112 unique domains receiving personal data from Android apps together with the personal data types received, obtained in our experiment.

# 2 Background and related work

## 2.1 Background

Identifying an organization receiving personal data requires a method capable of matching the receiver domain to the organization holding it. This section analyzes different techniques providing the necessary technical knowledge to comprehend our proposal.

A domain on the Internet is an authority that controls its own resources (e.g., a network or an IP address), and a domain name is a way to address these resources. Domain names are based on a hierarchy where Top Level Domains (TLD) represent the highest level (e.g., .org, .com, or .es) followed by Second Level Domains (SLD) (e.g., mozilla, google, or amazon). SLDs are managed by companies (i.e., domain name registrars) who register the information on authorities holding domain names in a global registry database. An authority can create subdomains to delimit areas or resources under its own domain (e.g., aws.amazon.com or www.amazon.com). A Fully Qualified Domain Name (FQDN), also known as an absolute domain name, is a domain name that specifies its exact location in the domain hierarchy.

WHOIS [11] is the standard protocol for retrieving information about registered domains and their registrants, including the domain holder's identity, contact details, domain expiration date, etc. Nevertheless, several issues [9] have been reported, including inconsistencies and lack of integrity in registrants' identity information.

Previous research has used WHOIS information for different purposes e.g., to extract registration patterns in the com TLD [12], to categorize organizations that own Autonomous Systems on the Internet [13], or to identify domains that redirect to malicious websites [14]. However, Watters et al. [15] pointed out that the basic deficiency in WHOIS data is a lack of consistency and integrity in the registrants' identity data. This was backed by an ICANN report [16] recognizing extended accuracy failures with only 23% of WHOIS records with 'No failure'. Aiming to address these concerns, the ICANN created the Accuracy Reporting System project [17], whose third phase (i.e. the one addressing registrant identity details) is in a to-be-defined status. Recent studies (e.g. [13]) still report that registrars inconsistently collect, release, and update basic information about domain registrants.

An SSL certificate can be another source of information about the authority holding a domain as it digitally binds a cryptographic key, a domain, and, sometimes, the domain holder's details. The cryptographic key allows for setting up secure connections (HTTPS) between the server and any requesting client.

Thus, whenever an HTTPS connection is set, the client can analyze the certificate used to get information on the server domain holder. HTTPS connections have grown over time, reaching 95% of the total connections in November 2022 [18].

SSL certificates are usually issued by a Certificate Authority (CA), which checks the right of the applicant organization to use a specific domain name and may check some other details depending on the certificate type issued. Extended Validation (EV) certificates are issued after the CA conducts a thorough vetting of the applicant and include information on their legal identity. Organization Validated (OV) certificates are issued after the CA conducts some vetting of the applicant and include information about the applicant's Organization Name under the ON field. Domain Validated (DV) certificates are issued with no vetting of the organization's identity, and no information about the applicant is included. Some studies report that DV certificates account for around 70% of all certificates [19].

An alternative technique to identify a domain holder is to search for it in the publicly available privacy policy that should be associated with that domain on the internet. A privacy policy, also known as a privacy notice, is typically presented as a textual document [43] through which an organization informs its users about the operations of their personal data (e.g., collection and transfer) and how it applies data protection principles. In many jurisdictions, e.g., the European Economic Area (EEA), the United Kingdom, or China, the privacy policy must also include the identity and the contact details of the personal information handler, or first-party or data controller in data protection parlance. It is reasonable to assume that the data controller for a given domain is also the authority holding that domain and vice versa.

## 2.2 Related work

WHOIS and SSL certificates are legit ways of identifying a domain holder. However, given the problems shown in the previous section, we have had to resort to a new method based on identifying the domain's owner through its privacy policy.

A set of activities are mainly needed to achieve this goal, primarily finding and analyzing the policy, and previous works have partially addressed them. For example, PolicyXray [10] tries to find the privacy policy for a specific URL by crawling all possible resources under that domain. Our method improves PolicyXray by considering other means to find the privacy policy associated with a domain, such as keywords (e.g., privacy, legal) search on the domain's home page and through external search engines (i.e., Google). Furthermore, ROI limits the number of requests to the domain to five, thus outperforming PolicyXray, since crawling a whole domain usually requires hundreds of requests that can overload the domain server [20].

Once a privacy policy has been found, the information identifying the data controller needs to be extracted from the text.

Del Alamo et al. [21] have provided an extensive review of the available techniques for the automated analysis of privacy policies and the information extracted from them. over symbolic ones for extracting data controller information, and within them. According to this survey, statistical Natural Language Processing (NLP) techniques are favored supervised machine learning algorithms are mostly reported.

Supervised learning algorithms are usually employed to select the policy segments (roughly speaking, a paragraph) where the content of interest is to be found [22]. They need a labeled (annotated) dataset to learn a specific characteristic of the text they will select. Although different authors have proposed techniques for crowdsourcing annotations of privacy policies (e.g., [23]), researchers' annotations supported by legal experts' assistance (e.g., [2]) are easier to collect for small datasets.

The techniques above show a good performance for classification problems i.e., determining the presence/absence of specific information in a privacy policy. For example, Torre et al. [24] followed this approach to determine the presence of a data controller's identity in a privacy policy. Costante et al. [25] applied it to understand whether the policy disclosed the data controller's contact details, e.g., postal address or phone number. Unfortunately, none extracted the controller's identity, just determining whether or not it was disclosed.

However, we aim to find and extract an organization's identity, and Named-Entity Recognition (NER) techniques are usually applied. Closer to our work, Hosseini et al. [26] used NER techniques to identify third-party entities on privacy policies. They trained three NER models with different word embeddings to obtain their results. This work differs from ours as their goal was to recognize all entities of a class (i.e., organization) in a policy. Instead, we aim to get only one output (the data controller identity) from all possible organizations (i.e., first party, third parties) disclosed in the policy text.

Analogously to our work, WebXray [27] also provides information about the holder of a given domain by combining WHOIS information with other information available on the web (e.g., Wikipedia). Several authors [28–30] have leveraged WebXray to identify organizations receiving personal data flows. Therefore, WebXray is the closest approach to compare our results with, which we do in section 3.4.

# 3 Method

WHOIS service consultation, SSL certificates inspection, and privacy policies analysis are three different techniques to obtain information on an organization receiving personal data. We detail our approach to extracting information from the WHOIS service and privacy policies below, together with their evaluation results. Finally, we propose and evaluate ROI, a new method that combines the techniques showing the best performance.

## 3.1 WHOIS consultation

We have followed two different approaches to query and parse the WHOIS records. First, leveraging a well-known Python library that queries and parses different WHOIS records. Second, developing our own module focused on extracting the registrant details.

For the first approach, WHOIS domain registration information was retrieved using the python-whois library [31], with over eight million downloads (over 130,000 in October 2022). After an in-depth analysis of the information recovered, we observed incomplete or missing fields that were not correctly parsed, particularly those related to the Registrant Organization identity. This is probably due to the absence of a consistent schema during the domain registration process, as noted by previous research [12]. Thus, we developed our own code to query the WHOIS service using the command line tool and parse the Registrant Organization details. We applied a final filter to discard values hidden for privacy reasons. This filter detects keywords, e.g., "redacted" or "privacy", from a bag of words.

## 3.2 Privacy policy analyzer

This technique departs from an absolute domain name, finds the privacy policy governing that domain, and analyzes it to extract the data controller identity (Fig. 1).

Finding the privacy policy governing a domain is not straightforward, as we depart from an absolute domain name and must get the website's home page URL to start the search process. To this end, we first obtain the SLD from the domain and then send an HTTP request to it, aiming to be redirected to a valid resource. In case of failure[1], we leverage search engines (i.e., Google) to find the home page for the given SLD. Once the home page is found, we search for the privacy policy. Again, we have followed two different approaches 1) Scraping the home page with Selenium to find the link to the privacy policy and, in case a valid policy is not found, and 2) searching the policy on the Google search engine.

When a potential privacy policy is found, its text is downloaded and kept for further analysis. Previous work has highlighted [32] that dynamic Javascript code is sometimes used to display a privacy policy. We relied on Selenium [33] to retrieve the complete text of the privacy policies, which deals with dynamic Javascript code. In our experimental tests, these techniques correctly found 65% of the privacy policies governing the target domain.

Once a potential privacy policy is collected, its language is checked with the langdetect [34] python package, and non-English texts are discarded. Afterward, a supervised Machine Learning (ML) model based on Support Vector Machines

---

[1] Our tests with different methods showed that scraping the target website first and then searching the policy in Google if the scraping didn't work yielded the best results.

(SVM) checks whether the text is indeed a privacy policy.

We applied the SVM approach to determine the optimum separation hyperplane for dividing the analyzed texts into privacy policies or other texts. SVM has empirically proved superior performance in high-dimensional spaces over a range of other ML algorithms, remaining successful even when the number of dimensions exceeds the number of samples [44].

Prior work [22], [45] revealed that SVM outperforms Logistic Regression and Convolutional Neural Networks for categorization of privacy practices. Relying on our previous experience building these kinds of classifiers [46], the hyperparameters used are the *Modifier-Huber loss function* and *SVM alpha* of $10^{-3}$.

We trained the model with 195 manually classified texts, achieving 98.76% precision, 97.56% recall, and 98.15% F1 score[2] when evaluated against 100 unseen English texts.

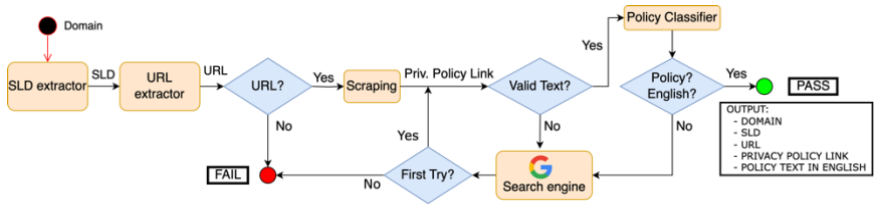

**Fig. 1** Privacy policy retrieval and analysis.

To identify the data controller in the privacy policy, we first select the paragraphs of the text where it is likely to appear. This selection is based on a bag of words that seeks keywords empirically demonstrated to be closer to the data controller disclosure (e.g., keywords such as "we" and "us" found in the TikTok app privacy notice as shown in Fig. 2). Following previous research in privacy policies analysis [22][45], our initial approach to identify paragraphs containing controller details was based on a machine learning SVM model trained with 100 manually annotated privacy policies. Nonetheless, privacy policies typically follow a common format and structure. Specifically, the consistent structure of paragraphs where the policy's data controller is declared has led to better results using alternative techniques such as keyword search (i.e., Bag of Words), which was finally implemented.

```
Welcome to TikTok (the "Platform"). The Platform is provided
and controlled by TikTok Inc. ("TikTok", "we" or "us"). We
are committed to protecting and respecting your privacy. This
Privacy Policy covers the experience we provide for users age
13 and over on our Platform.
```

**Fig. 2** Privacy policy example (TikTok app) showing the data controller disclosure.

---

[2] Accuracy, precision, recall, and F1 score are measures used in Machine Learning to evaluate the performance of categorization algorithms. Accuracy is calculated by dividing the number of correctly classified cases by the total number of instances. Precision is the proportion of true positives among all positive instances, whereas recall is the proportion of true positives among those cases that genuinely belong to the positive class. The F1 score is the harmonic mean of precision and recall, providing a balance between the two measurements.

Named Entity Recognition (NER) techniques are applied to the selected paragraphs to identify the data controller. We have used SpaCy [35] for this, which provides two different trained NER models, one prioritizing efficiency and another favoring accuracy. After testing both, the efficiency model showed poorer results, so the accuracy-based model was implemented. We assessed the performance of the combination of the bag of words and the NER with 142 privacy policies, obtaining the results shown in Table 1.

**Table 1** Privacy policy analyzer metrics.

|  | Accuracy | Precision | Recall | F1-score |
|---|---|---|---|---|
| Data controller extraction | 92.25% | 95.45% | 94.59% | 95.02% |

## 3.3 Individual techniques evaluation

The main goal of this study is to identify the organizations receiving flows of personal data. These organizations can present substantial differences, e.g., company size, location, etc. To make our evaluation as fair as possible, we used a subset of 100 domains randomly chosen from a larger set of 1,004 domains that we found receiving personal data from a previous experiment we carried out [2]. For each domain, we manually searched the privacy policy and the data controller disclosed herein.

We used this random dataset to evaluate the performance of WHOIS service consultation, SSL certificates inspection, and privacy policies analysis to identify the controller behind a given domain (Table 2). The output of each technique is either 1) a given value for the domain holder, which can be right (i.e., true positive - TP) or wrong (i.e., false positive - FP), or 2) no value (i.e., false negative - FN) in case the technique cannot determine a specific domain holder. A result cannot be considered as true negative as every domain must have a holder, even if a technique is not able to find it.

The inspection of SSL certificates found 99 certificates out of 100 domains fed, with 30 of them containing the organization name. The missing certificate could not be obtained because this domain uses an HTTP connection. Twenty of the organization names retrieved were correct and ten were wrong; the remaining 69 certificates did not contain the organization name. We did not find any kind of relation between the identity of the CA issuer and the absence of the organization's identity in the certificates. Only two of those certificates contained IP addresses in addition to the CN. These results translate into a 66.67% precision score, but only 20% identified organizations.

We evaluated the python-whois library as well as our own implementation. Python-whois couldn't find information for ten domains. From the remaining 90 registries, 34 were obtained, 30 were hidden for privacy reasons, and 26 were not correctly parsed. Our own WHOIS-parsing implementation obtained 37 valid owners and 2 incorrect ones. For the remaining 61 domains, 24 did not contain the Registrant Organization field, 2 had an empty value on this field, and 35 registries were hidden for privacy reasons. These results entail the python-whois library performed an 87.18% precision score while our implementation scored 94.87% precision, obtaining more correct results.

Finally, our privacy policy analyzer was evaluated, achieving the best results of all tested techniques. The evaluation is applied to the whole pipeline, including extracting the privacy policy associated with the targeted domain and extracting the data controller name. This pipeline is therefore affected by the performance of each step. Nevertheless, its results outperformed the other techniques, with 56 correct and 4 incorrect outputs, meaning a 93.34% precision score.

**Table 2**   Comparative between all the individual techniques.

|  | Accuracy | Precision | Recall | F1-score |
|---|---|---|---|---|
| SSL certificate inspection | 20.00% | 66.67% | 22.23% | 33.34% |
| WHOIS consultation | 30.00% | 94.87% | 37.75% | 54.01% |
| Privacy policy analyzer | 56.00% | 93.34% | 58.33% | 71.79% |

## 3.4 ROI: Receiver Organization Identifier

Given the results achieved by the individual techniques and after a detailed analysis, we combined the privacy policy analyzer with the WHOIS consultation into ROI, a new method to identify an organization receiving personal data. The SSL certificate inspection was discarded due to their low precision score. The python-whois library was also discarded in favor of our implementation.

Interestingly, the combination of the privacy policy analyzer as the first choice and our WHOIS implementation as the second choice outputs the best results, showing even better precision score (95.71%) as the individual techniques while considerably reducing the number of false negative results, achieving 67 true positive results with only 3 false positive results. ROI operating scheme is represented in Fig. 3.

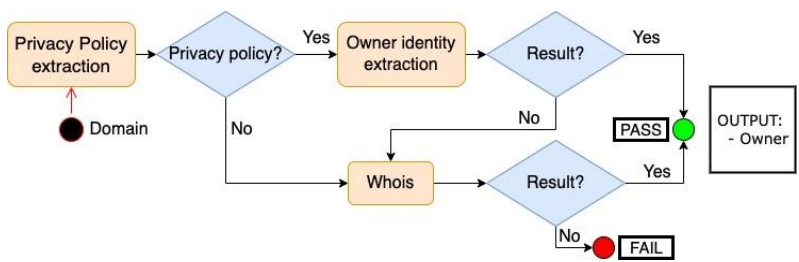

**Fig. 3**   ROI operation flow diagram.

We further analyzed the three false positives. Our NER failed to identify the data controller in two of them. Interestingly, in one of the cases the privacy policy did not mention any data controller at all, which goes against the transparency requirements set by GDPR. As for the third false positive case (unseenreport.com), ROI wrongly attributed this domain to Google as our HTTP request to unseenreport.com was redirected to google.com, raising a red flag due to the redirection to a different SLD. Unfortunately, we were not able to find the holder organization, even after carrying out an in-depth search of this domain. The domain has been categorized as a malicious website by ANY.RUN [36].

We did a manual inspection on the 30 domains that ROI could not identify. Eleven of these domains did not provide a landing page and are probably only used for back-end purposes.

We did find a landing page for other 13 domains but could not find their privacy policy even while receiving personal data, which may raise compliance issues according to GDPR. Three domains had a non-English website, which is a ROI limitation analyzed in section 5. From the remaining three, one provided the privacy policy through JavaScript expanding elements which were not automatically triggered. These results prove the good performance of ROI, showing that non-identified organizations are usually not providing the information mandated by privacy regulations and thus deserving closer inspection, opening new research lines.

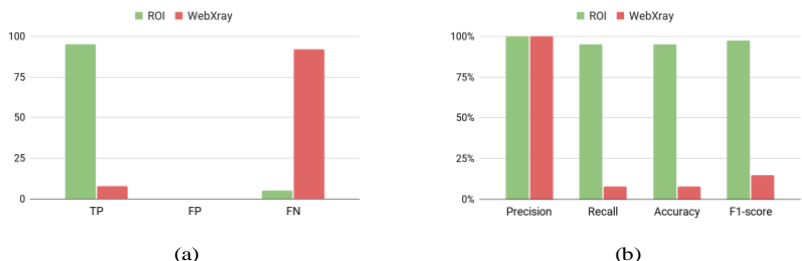

**Fig. 4** Comparison between ROI & WebXray results (a) and metrics (b) on the Fortune-500 dataset.

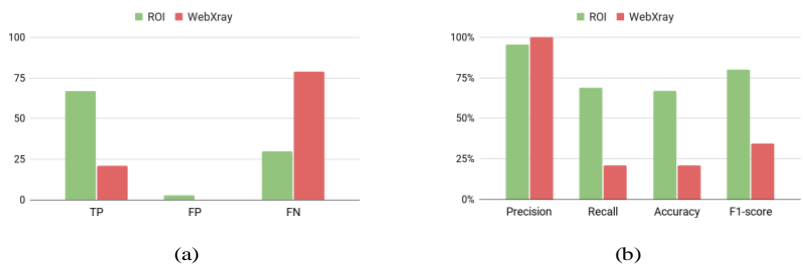

**Fig. 5** Comparison between ROI & WebXray results (a) and metrics (b) on a random sample of personal data receivers.

Like ROI, the WebXray tool identifies the organization behind a given domain receiving personal data [27]. Thus, we have compared ROI and WebXray performance against two unseen ground-truth datasets. The first dataset includes 100 URLs (homepages) held by Fortune-500 companies. The second dataset, similar to the one described in section 3.3, consists of a random sample of 100 URLs receiving personal data flows.

Fig. 4 illustrates the comparison between the True Positive, False Positive,

and False Negative cases (a) and the precision, recall, accuracy, and F1-score metrics (b) from the evaluation of the Fortune-500 dataset. The same comparison is performed on the random sample of personal data receivers (Fig. 5).

There is a noticeable difference in ROI's performance against each dataset. Indeed, ROI achieves slightly better results (97.44% F1-score) against well-known domains than against other less-known receivers (80.24% F1-score). This difference is due to the performance of the WHOIS consultation for the latter since the privacy policy analyzer behaved similarly in both cases.

ROI is a Python-based tool running in a Docker container with all its dependencies installed, so it can be easily deployed in new settings. However, searches are carried out through the Google official paid API and, therefore, the use of ROI requires a Google API token for each new deployment. Currently, ROI is able to process privacy policies written in English but not in other languages (only 7.34% of the policies we found in our experiments are not written in English).

All in all, ROI outperforms existing tools serving the same purpose, and its notable scalability and its low number of false negative results support applying it in identifying personal data receivers in the wild.

ROI can easily serve various stakeholders. For instance, researchers in fields such as data protection and privacy can leverage ROI in their research e.g. to uncover companies collecting massive amounts of personal data (as shown in section 4.2). In turn, data protection authorities can utilize ROI to assess mobile applications at scale and discover those sharing data with third parties without declaring it in their privacy policies (as demonstrated in section 4.3). Similarly, it can assist developers in correctly identifying and declaring these recipients in their privacy policies, mitigating substantial fines for non-compliance.

# 4 Demonstration: Android apps evaluation in the wild

In this section, we demonstrate ROI by evaluating 10,000 Android apps from the Google Play Store, analyzing what personal data they send out and the organizations receiving them, and checking whether the recipients have been properly disclosed in the apps' privacy policies. To this end, we describe our experimental environment, and report and analyze the results obtained.

## 4.1 Experiment setup

We developed a controlled experiment leveraging our previous work [2] on personal data flow interception and analysis in Android apps. This is a pipelined microservices-based platform made up of different modules able to automatically 1) search, download, install, run, and interact with Android apps, and 2) intercept and analyze outgoing network connections.

Specifically, the Download module logs into the Google Play Store simulating a real device and downloads the applications, storing them in the Storage module. The Traffic module is a multi-threaded Python script handling multiple real devices connected at the same time. It gets applications from the Storage module and installs them in each device through the Android Debug Bridge (ADB) [48] connection. After the installation, it runs the apps first in an idle phase (without app stimulation) and then in a dynamic phase (with automated stimulation using Android Monkey [47]). At the same time, a Man-in-the-Middle proxy and an application instrumentation tool (Frida) are used to intercept and decrypt secured connections. The connections' payloads are decoded trying different formats (e.g., Base64, SHA, MD5) and inspected looking for personal data. The results are logged to our centralized logging platform based on ElasticSearch.

Previous research has extensively addressed the detection of personal data leaks in Android apps following two approaches, namely static and dynamic analysis. Static techniques [37] focus on detecting data leakages by analyzing the code without executing it. On the other hand, dynamic techniques require the apps' execution and a further interception of the communications either inside the device (e.g., setting up a virtual private network and analyzing the outgoing traffic [38]). Our setup favors dynamic analysis techniques to capture network packets generated by the real execution of apps rather than static analysis techniques that analyze approximate models that, while ensuring high recall, could generate a high rate of false positives.

Our platform was fed with a list of 10,000 random Google Play Store apps from the top-downloaded category. The apps were collected, downloaded, installed, and executed in September 2022 on five mobile devices Xiaomi Redmi 10, running Android 11 (API 30). Following common practices for dynamic analysis in Android [39], the idle phase was performed for two minutes and the Android Monkey was used to interact with each application for an extra three minutes. Considering the five devices running uninterruptedly and ignoring devices' bugs (which forces us to manually restart the device affected) it required 6 days 6h and 35 minutes to finish our analysis.

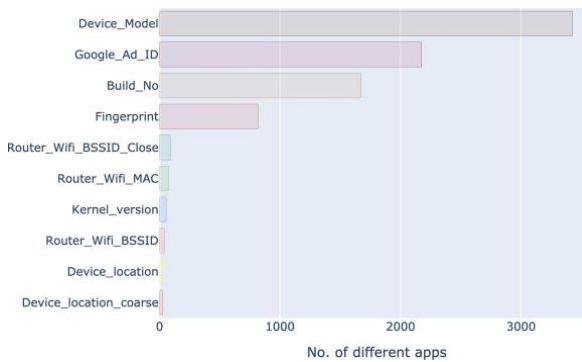

(a)

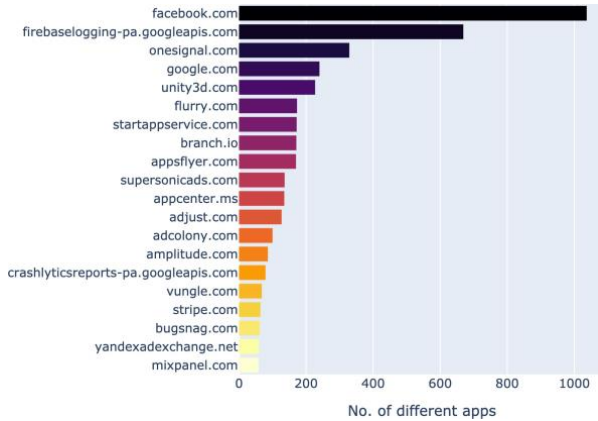

(b)

**Fig. 6** Personal data sent off the device (a) and the popular destinations (b).

## 4.2 Recipients' analysis

Our platform managed to execute 7,037 apps, identifying 40,493 personal data flows from 3,526 apps to 1,112 unique domains during the experiment. A vast portion (99.2%) of these data flows correspond to HTTPS connections, which are aligned with the HTTPS encryption level observed on the Web [18]. Interestingly, we found 320 (0.8%) HTTP connections containing personal data, which is an insecure practice. Alarmingly, these HTTP connections included all types of personal data we found except the device's software build number. Therefore, personal data such as the Google advertising identifier or the device location are being sent without adequate protection.

Fig. 6 shows the number of apps sending out each personal data type (top), and the number of apps that sent personal data to the top-20 destination SLDs (bottom). Interestingly, most apps sent out the device model name (97.13%), and more than half of apps (61.68%) sent the Google advertising identifier, which is closely aligned with what was observed by previous research [49].

Fig. 7 further details the types of personal data that the top-10 domains are receiving. We can see that nine out of ten domains are collecting the Google advertising identifier, commonly used for monetization tracking and personalized advertising. As could be expected, most of the top-20 domains receiving personal data are for analytics, marketing, or monitoring purposes (e.g., firebaselogging-pa.googleapis.com, supersonicads.com, adcolony.net).

| Type of personal data | Domains | | | | | | | | | | Domains |
|---|---|---|---|---|---|---|---|---|---|---|---|
| | 1 | 2 | 3 | 4 | 5 | 6 | 7 | 8 | 9 | 10 | |
| Device_Model | ✓ | ✓ | ✓ | ✓ | ✓ | ✓ | ✓ | ✓ | ✓ | ✓ | 1. facebook.com |
| Google_Ad_ID | ✓ | | ✓ | ✓ | ✓ | ✓ | ✓ | ✓ | ✓ | ✓ | 2. firebaselogging-pa.googleapis.com |
| Build_No | ✓ | ✓ | | | ✓ | ✓ | | ✓ | ✓ | ✓ | 3. onesignal.com |
| Fingerprint | ✓ | | | | | | | | | | 4. google.com |
| Router_Wifi_BSSID_Close | | | | | | | | ✓ | | | 5. unity3d.com |
| Router_Wifi_MAC | | | | | | | | ✓ | | | 6. flurry.com |
| Kernel_version | | | | | | | | | | | 7. startappservice.com |
| Router_Wifi_BSSID | | | | | | | | | | | 8. branch.io |
| Device_location | | | | | | | | | | | 9. appsflyer.com |
| Device_location_coarse | | | ✓ | | | ✓ | | | | | 10. supersonicads.com |

**Fig. 7** Type of personal data received by popular domains.

We further applied ROI to identify the companies holding the domains receiving the personal data. Overall, we determined them in 82.37% (33,356)  of the personal data flows, representing 68.7% (764) of the unique destination domains. Fig. 8 shows how many apps sent personal information to the top collectors.

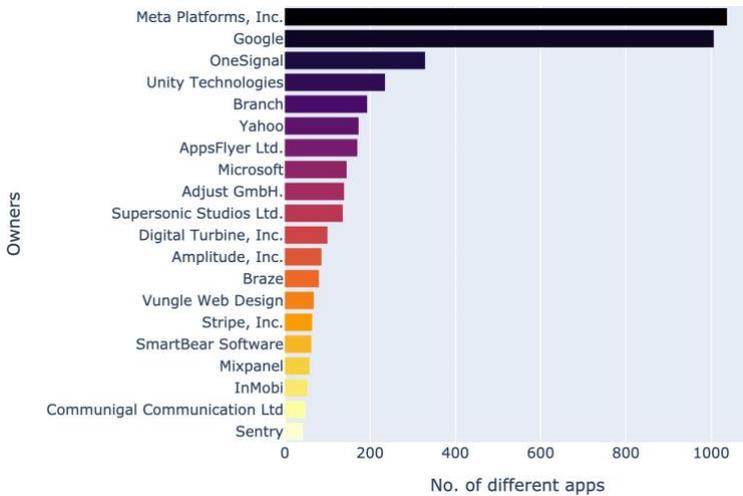

**Fig. 8** Companies receiving personal data.

The top-6 companies to which most apps send personal data provide analytics and marketing services. Furthermore, Meta and Google lead this list, receiving data from 1,037 (29.41%) and 1,006 (28.53%) apps, respectively. Indeed, half of the apps (51.56%) sent personal data to either Meta or Google.

Other companies from this top 10, e.g., Unity or Supersonic Studios Ltd., support games development and publishing, which means the importance of the gaming category in the Google Play store market. On a curious note, Sentry, which provides error and crash monitoring services, also receives users' personal data.

We further analyzed the hierarchy of relationships of the companies we found. The purpose is to make a representative illustration (Fig. 9) of the companies that collect the greatest volume of data, showing the head company as the representative. For example, Microsoft is the parent company of GitHub and LinkedIn. To achieve this, we employed the Crunchbase API [40] to enrich information about an organization, including the hierarchy of relationships between companies, e.g., parent and subsidiaries. Specifically, we searched in Crunchbase for the name of the company we found with ROI and obtained its country, description, and relationships with other companies. We repeated the process iteratively until we found the head company and all the companies under its control.

Fig. 9 demonstrates that some companies might receive data from several subsidiaries. Thus, the actual amount of personal data collected might be higher than expected, as for Fig. 8. The example of AppLovin is quite representative. While AppLovin provides users with monetization tools, they also have Adjust as a subsidiary for helping developers and MoPub for advertisement serving. The result is a whole ecosystem of companies collecting data that situate the corporation in the top-6 according to our data, above Verizon. Microsoft is another example of a company with several subsidiaries collecting data, i.e., LinkedIn and GitHub.

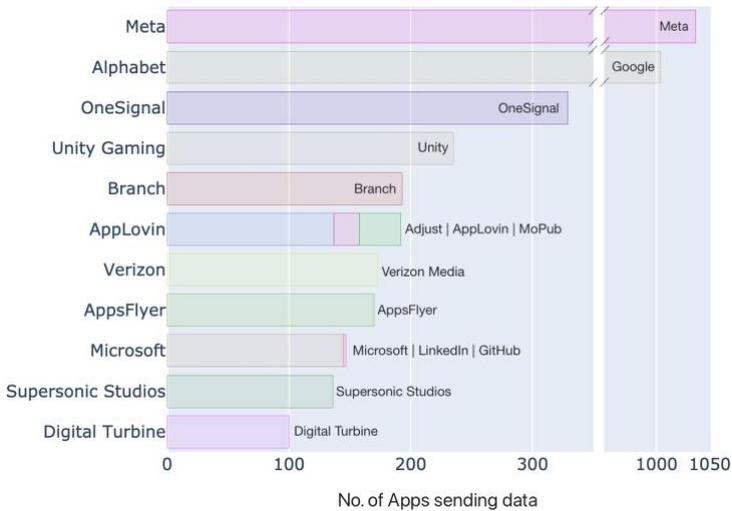

**Fig. 9** Top 10 head companies with subsidiaries receiving personal data.

Interestingly, the top 8 head companies provide well-known Software Development Kits (SDK) for Android, appearing on the Google Play SDK index [41]. Therefore, it is fair to assume that most personal data receivers are third-party organizations. This is a common practice in the mobile ecosystem and was addressed in previous research [28].

## 4.3 Recipients disclosure

ROI facilitates transparency and accountability in data protection practices. For example, valuable insight can be gained by cross-referencing the third-parties disclosed in privacy policies with those identified through ROI. When app developers and recipient organizations provide transparent information about the entities involved in data exchange, they demonstrate a commitment to accountability and foster user trust. Through improved transparency, users gain a deeper understanding of how their personal data is handled, empowering them to make informed decisions about apps.

To further demonstrate the ROI potential, we have checked if the recipients identified in the previous section are actually disclosed in the apps' privacy policies. To this end, we retrieved and processed the privacy policies of 2,155 applications, extracting the third-parties mentioned therein. We could not process 1,371 apps from our initial dataset as we could not find their privacy policy, which already flags a huge bunch of apps potentially non-compliant with applicable data protection laws such as GDPR.

Android apps generally fail to disclose their data-sharing practices. Only 476 (22%) apps accurately reflect all the entities receiving personal data in their privacy policies, while 1,327 (61%) do not disclose any of these entities. The remaining 352 applications fail to declare at least one recipient, providing only partial disclosure. This makes an outstanding 78% of the apps analyzed failing to fully disclose their data-sharing practices as mandated by data protection laws (e.g GDPR Art. 13 (1)(e)).

The lack of disclosure does not equally affect all personal data recipients. Figure 10 illustrates that Google is the recipient most frequently disclosed by the applications (464 out of 1,327 apps properly disclose the transfer of personal data to this company), significantly ahead of the others. Conversely, our observations show that applications very often fail to mention Meta in their privacy policies. A total of 595 applications out of the 828 that send data to Meta (consciously or unconsciously) fail to declare it.

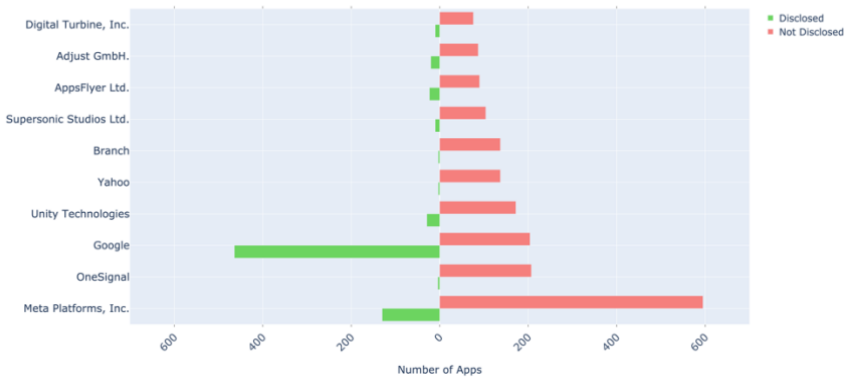

**Fig. 10** Third-party disclosure of top organizations in Android apps' privacy policies.

Some specific cases are alarming. For example, "com.wallapop", a leading second-hand selling application in Spain with 10 million downloads on the Google Play Store, fails to disclose recipients like Tapjoy and Google (among others), which are receiving the Google advertising ID and the device's fingerprint, respectively. Fortunately, we have also found full disclosures e.g. the "jigsaw.puzzle.free.games" application, which has 50 million downloads on the Google Play Store, accurately declares data transfers to Meta, Unity, Google,

Amazon, and other major tech entities.

Compliance with privacy and data protection laws is critical in today's regulatory world. As we have demonstrated, ROI can substantially aid in identifying personal data recipients, thus supporting compliance assessment processes.

# 5 Threats to research quality

ROI is a reliable and highly precise method to identify organizations holding domains receiving personal data and can identify the majority of the tested domains. Nevertheless, some limitations have been identified during the development process.

Finding the privacy policy for a given domain is at the core of ROI. However, the disclosure of privacy policies is not standardized; thus, many ad-hoc means are used to present the policy text, e.g., contained in popup elements. Following best practices in the field [23], we have relied on Selenium to address most of these issues as it deals with dynamic JavaScript code.

The information disclosed within a policy text (i.e., the data controller) is not standardized either. Although privacy policies are mandatory in some jurisdictions, e.g., the European Union as for its General Data Protection Regulation (GDPR) [42] Article 13, this information is often missing or wrong (e.g., frequently the app name is used to refer to the data controller, even if the app name is not a legally registered organization). We have addressed this challenge by partially validating the controller extraction method, achieving a nearly 95% F1 score, proving the good performance of this method.

Another limitation comes from the privacy policy text language. For the time being, our NER works with English texts, and it cannot process texts in other languages. Thus, we discarded non-English texts in our analysis, corresponding to only 7.34% of the policies found. To reduce the amount of non-English policy texts, we configured our tools to favor English texts. This was achieved by setting the accept-language parameter in the requests' headers and the lang argument in Selenium's configuration. Nevertheless, we are working on translation methods with NLP techniques that will help us to improve the number of privacy policies analyzed.

Our experiment involved results from 3,526 apps. The results with this amount of apps are representative, but outliers may appear when speaking about some data receivers. The results can be extended to consider a more significant number of applications, thus supporting the generalization of the results.

Finally, automated access to web pages might be viewed as unethical if it overloads the website. However, ROI only makes a maximum of five requests per domain instead of crawling them with hundreds of queries. This was possible thanks to the bag of words technique (cf. section 3.2 for details) we applied.

# 6  Conclusion

This paper has described ROI, a new method that leverages the information available in privacy policies and the WHOIS service to identify organizations receiving personal data flows. ROI achieves a 95.71% precision, greatly outperforming similar methods in the state of the art. We have demonstrated its applicability in the Android context by identifying the companies receiving personal data from 3,526 apps in the wild. Unfortunately, we have also shown that a huge portion of these apps fail to properly disclose these organizations in their privacy policies.

ROI brings benefits to various stakeholders. Data protection authorities can leverage it to understand the compliance of personal data collecting systems with privacy and data protection regulations. App developers can gain valuable insights into how their applications adhere to them. Researchers can gain a better understanding of the destinations of massive amounts of personal data.

Our future work points towards contributing to new techniques that support the privacy engineering community in automating the assessment processes of digital systems and services. To this end, we are leveraging ML and NLP techniques to automate the extraction of transparency elements from the privacy policies and check them against the actual behavior observed in the systems under analysis.

## Acknowledgments

This work has been partially supported by the TED2021-130455A-I00 project funded by MCIN/AEI/10.13039/501100011033 and by the European Union "NextGenerationEU"/PRTR; and, by the Comunidad de Madrid and Universidad Politécnica de Madrid through the V-PRICIT Research Programme 'Apoyo a la realización de Proyectos de I+D para jóvenes investigadores UPM-CAM', under Grant APOYO-JOVENES-QINIM8-72-PKGQ0J. It was possible to identify the relationships between parent and subsidiary companies thanks to Crunchbase, who kindly allowed us free access to its API for this research.

## Conflict of interest

The authors declare through the submission of this document that they have no conflicts of interest. This work was partially supported by the European Union, the Comunidad de Madrid. These are public funds granting that there are no financial or personal interests related to the research. The authors received no financial support or other benefits from any organization or individual with a stake in the research.